\documentclass{optica-article}

\journal{opticajournal} 

\articletype{Research Article}

\usepackage{amsmath}
\usepackage{lineno}
\usepackage{mathtools}
\usepackage[section]{placeins}
\usepackage{tabularx}
\usepackage{booktabs} 

\begin{document}

\pagenumbering{arabic}
\pagestyle{plain}

\title{A thermodynamic approach to linear cross-talk in multimode fiber systems}

\author{Mario Zitelli\authormark{1*}}

\address{\authormark{1}Department of Information Engineering, Electronics and Telecommunications (DIET), 
Università degli Studi di Roma Sapienza, Via Eudossiana 18, Rome, 00184, RM, Italy}
\email{\authormark{*}mario.zitelli@uniroma1.it} 


\begin{abstract*} 
Optical thermodynamic theory is extended to low-power multimode fiber systems to characterize with simple thermodynamic models the complex scenario of power exchange induced by random mode coupling between propagating modes. It is theoretically and experimentally demonstrated that low-power multimodal systems can give rise to steady-states described by a weighted Bose-Einstein modal power distribution.
The theory applies also at quantum power level, indicating that optical thermodynamics and quantum optics coexist, allowing the study of multimodal optical systems in both classical and quantum regimes and simplifying the design of SDM systems.

\end{abstract*}

\section{Introduction}

Multimode (MMF) \cite{Gloge:6767859,Hasegawa:80} and multicore (MCF) \cite{Inao:79,Noordegraaf:09} optical fibers have been the subject of intense studies in the last decade, in order to solve the problem of capacity crunch in optical networks \cite{Essiambre:5420239,Winzer:18} using the space-division multiplexing (SDM) technique \cite{Richardson-NatPhot-2013-94-2013}, which adds a further degree of freedom for increasing the transmission capacity by multiplexing channels into the different orthonormal modes or cores of a fiber. In fact, SDM is compatible and cumulative with other multiplexing technologies, such as the wavelength-division multiplexing (WDM), the polarization and time multiplexing. 

Both MMFs \cite{Wang:20,Zhong:21} and MCFs \cite{Dynes:16,agrawal_2023} have been tested for use in quantum-key distribution (QKD) applications, in order to multiply the qubit rate aggregated capacity, or to multiplex into different fiber modes or cores both quantum and classical signals. However, not enough research has been carried out to maximize the impact of such an application. 

The main transmission impairments of MCF and MMF low-power systems are the modal cross-talk, consisting of a random exchange of power between orthonormal modes of a fiber, and the mode-dependent losses (MDL) which are different from mode to mode and strongly affect the system performance \cite{Gloge:6767859}. The physical effect causing modal cross-talk is the random-mode coupling (RMC) \cite{Gloge:6774107,Savovi2019PowerFI,Ho:14}, caused by fiber imperfections such as micro or macro bending and chaotic variations in multimode fiber diameter.

Multiple-input multiple-output (MIMO) signal processing is a computational technique used at the receiver able to recover the cross-talk between channels \cite{Winzer:11,Randel:11}; it has also been used in QKD wireless communications \cite{Gabay:06}. In optical fiber SDM systems, it consists of a linear combination of the information transmitted to the receiver by the $M$ different modes or cores of the fiber, in order to reconstruct the transmitted channels and compensate for the modal cross-talk. Linear combinations require $M x M$ arrays of coefficients which are dynamically refreshed; for this reason, a massive use of resources may be required to estimate the coefficients.
If the transmitter is able to address all the propagation modes supported by the fiber, and if the receiver is capable of mode-selectively detect all propagation modes, MIMO technologies are able to recover the modal cross-talk in MCFs and MMFs, allowing designers to increase up to $M$ times the number of modal channels to be used for both quantum and classical information transmission, thus reducing the impact of quantum channels on the overall transmission capacity of a fiber system. 

In this context, optical thermodynamics \cite{wu2019thermodynamic,Fusaro_PhysRevLett.122.123902,pourbeyram2022direct,PhysRevX.14.021020} appears to be a useful tool for simplifying the design of SDM systems, provided its validity is demonstrated for low-power and quantum transmission systems. The use of thermodynamic laws permits to characterize the modal power distribution at the output of a MMF or MCF in terms of two or three parameters; once the output distribution is known, the development of RMC mitigation techniques, such as the MIMO algorithms, can be eventually simplified by imposing constraints.

In this work, the optical thermodynamic theory will be extended to low-power linear systems using graded-index multimode fiber (GRIN MMF), where the power exchange between modes is to be attributed to the RMC rather than to nonlinear processes such as the inter-modal four-wave mixing (IM-FWM) \cite{wu2019thermodynamic,zitelli2023statistics}. It will be theoretically and experimentally shown that RMC in low-power multimodal systems can give rise to steady-states, in which power distribution between system modes evolves irreversibly towards the thermodynamic equilibrium state of maximum Boltzmann entropy, corresponding to a weighted Bose-Einstein (wBE) modal power distribution. 
Experiments will be performed using both low-power pulses or continuous-wave signals at telecom wavelength, transmitted on a SDM systems composed by modal multiplexers/demultiplexers and up to 5 km of GRIN fiber. Experiments will be repeated using single-photon pulses in order to emulate a SDM quantum system. It will be shown that RMC acts between different modal groups as well among quasi-degenerate modes. In both linear and quantum regimes, the validity of the thermodynamic predictions to the experimental data will be demonstrated.

\section{Theory}\label{sec:Theory}

A GRIN MMF, and more in general a multimode system, is able to propagate $Q$ groups of degenerate modes, being $g_j$ the degeneracy over two polarizations of modal group $j$. In the specific case of GRIN fibers, it is $g_j=2, 4, 6,.., 2Q$ for $j=1, 2, .., Q$, respectively, and fiber supports $g_j/2$ modes per group; the total number of modes and polarizations is $2M=Q(Q+1)$. 


We suppose that $n_j$ is the photon population into modal group $j$, distributed over $g_j$ nearly-degenerate modes and polarizations. $\epsilon_j=\beta_j-\beta_{j=Q}$ are the differential modal eigenvalues, with $\beta_j$ the propagation constant in group $j$. The total number of photons in the system is $N=\sum_{j=1}^Q n_j$. 

In nonlinear systems, where IM-FWM is responsible for power exchange between modes \cite{Poletti:08}, it is convenient to relate $N$ to the peak power $P$ of the propagating pulse used in the experiments; we define in this case the internal energy as $U=-\sum_{j=1}^Q \beta_j n_j P_0/n_0$ (in units of W/m), with $P_0$ the peak power corresponding to a reference number of photons $n_0$ \cite{zitelli2023statistics}. Chromatic dispersion and other dissipative effects can invalidate the thermodynamic approach because they reduce the pulse peak power and, correspondingly, the number of distributed photons $N$ and the effects of IM-FWM.

In linear systems, RMC is the responsible of power exchange between degenerate and non-degenerate modes, which is properly described by the power-flow model \cite{Gloge:6774107,Savovi2019PowerFI}. The effect does not depend on the pulse peak power and still holds after an important pulse broadening, provided the pulsewidth is much larger than the modal dispersion-induced delay. In this case, $P$ is replaced by the mean power; alternatively, the number of photons is conveniently related to the energy $E_p$ of the propagating pulse, and the internal energy can be defined as $U=-\sum_{j=1}^Q \beta_j n_j E_{p0}/n_0$ (J/m), and $\gamma=N/n_0=E_p/E_{p0}$ is a fractional total number of photons.


The multiplicity of the system, as the number of possible microstates populated by the photons, is given by

\begin{equation}
W=\prod_{j=1}^Q \frac{(n_j+g_j-1)!}{n_j!(g_j-1)!}   .
\label{eq:Multiplicity}
\end{equation}

The Boltzmann entropy of the system $S=\ln(W)$ is related to the number of microstates; it reads for $n_j >> g_j$ to \cite{wu2019thermodynamic,zitelli2023statistics}

\begin{equation}
S=\sum_{j=1}^{Q}(g_j-1)\ln(n_j) .
\label{eq:EntropyB}
\end{equation}


The quantity $S_N=S/\gamma=n_0\ln{(W)}/N$ has the meaning of an entropy per unit particle; although it cannot be considered as an entropy itself (for example, it is not additive), we may try to find an extremum of $S_N$ while assuming constant the system's normalized internal energy $U_N=U/E_p=-\sum_{j}\beta_j n_j/N$ (1/m)
 
 \begin{equation}
\frac{\partial}{\partial n_l}\Big[\ln{(W)}/N+ \sum_{j=1}^Q \Big(a n_j/N +b \beta_j n_j/N \Big)\Big]=0      .
\label{eq:EntropyDerivative1}
\end{equation}

 Eq. \ref{eq:EntropyDerivative1} is multiplied by $N$ and solved using two possible sets of Lagrange multipliers $(a, b)$ or $(a', b')$ \cite{zitelli2023statistics}. 

 A first solution is related to the choice of non-factorizable constants $a$ and $b$ defined as

\begin{equation}
-(a+b \beta_j)=\ln{\Big[\frac{1}{n_0}\exp{\Big(-\frac{\mu+\beta_j}{T}\Big)}-\frac{1}{n_0}+1\Big]} \simeq \frac{1}{n_0} \Big[\exp{\Big(-\frac{\mu+\beta_j}{T}\Big)}-1 \Big] ,
\label{eq:DerivativeRes3}
\end{equation}

 
 which brings to the weighted Bose-Einstein law (wBE) 
 
\begin{equation}
\lvert f_j \rvert ^2=\frac{2(g_j-1)}{g_j\gamma}\frac{1}{\exp\big(-\frac{\mu'+\epsilon_j}{T}\big)-1}    ;
\label{eq:BE}
\end{equation}

in Eq. \ref{eq:BE}, $\lvert f_j \rvert^2=2 n_j/(\gamma n_0 g_j)$ is the mean modal power fraction, over two polarizations, in modal group $j$. $\mu'=\mu+\beta_{j=Q}$, with $\mu$ (1/m) a chemical potential and $T$ (1/m) an optical temperature. 
$\mu'$ and $T$ are two degrees of freedom for fitting Eq. \ref{eq:BE} to the experimental data. The $\gamma$ parameter is free at only one intermediate pulse energy; for other energies, it must scale proportionally to $E_p$ or $N$; the second constraint is the respect of the conservation law $\sum_{j=1}^{Q}(g_j/2)\lvert f_j \rvert ^2=1$.

Hence, the wBE distribution is valid in experiments where $U_N=const$ and when the total number of photons $N$ is conserved during transmission.

The power fraction $\lvert f_j \rvert^2$ is averaged into a modal group $j$ assuming statistical modal power equipartition into groups; in fact, RMC is responsible for a fast power exchange among quasi-degenerate modes.

A second solution, related to the choice $a'=\mu/(Tn_0)$, $b'=1/(Tn_0)$ provides an alternative Bose-Einstein solution

\begin{equation}
\lvert f_j \rvert ^2=\frac{2(g_j-1)}{g_j\gamma n_0}\frac{1}{\exp\big(-\frac{\mu'+\epsilon_j}{Tn_0}\big)-1}    .
\label{eq:BE_n0}
\end{equation}

Under the assumption $\lvert \mu'+\epsilon_i \rvert << \lvert T n_0 \rvert$, Eq. \ref{eq:BE_n0} leads to the well-known Rayleigh-Jeans (RJ) distribution:

\begin{equation}
\lvert f_j \rvert ^2=-\frac{2(g_j-1)}{g_j \gamma}\frac{T}{\mu'+\epsilon_j} \simeq -\frac{T'}{\mu'+\epsilon_j}    ,
\label{eq:RJ}
\end{equation}

with $T'=2T/\gamma$ (1/m). 
The two sets of Lagrange multipliers used to obtain Eqs. \ref{eq:BE} and \ref{eq:BE_n0} bring to the same values of internal energy and number of photons, namely \cite{zitelli2023statistics}

\begin{equation}
\sum_{j=1}^Q (a + b \beta_j) n_j=\sum_{j=1}^Q a' n_j + b' \beta_j n_j=Q-2M= \frac{\mu N}{T n_0}+\frac{1}{T n_0}\Big(-\frac{UN}{E_p}\Big)  ,
\label{eq:StateEquation1}
\end{equation}

\noindent which provides a common state equation (SE)

\begin{equation}
U-\mu E_p=(2M-Q)TE_{p0}.
\label{eq:StateEquation2}
\end{equation}

The SE can be rewritten in terms of $U_N=U/E_p$, $\gamma$ and fitting parameters $\mu'$ and $T$

\begin{equation}
SE=-U_N+\mu+V\frac{T}{\gamma}=\sum_{j=1}^Q \beta_j \frac{g_j}{2}\lvert f_j \rvert^2 +\mu'-\beta_{j=Q} +V \frac{T}{\gamma}=0 ,
\label{eq:StateEquation3}
\end{equation}

where we introduced $V=(2M-Q)$ as the system volume. The experimental error on the SE can be calculated as

\begin{equation}
\epsilon_{SE}=\frac{SE}{\sum_{j=1}^Q \beta_j \frac{g_j}{2} \lvert f_j \rvert^2 -\Big(\mu'-\beta_{j=Q} +V \frac{T}{\gamma}\Big)} .
\label{eq:StateEquationError}
\end{equation}

The error $\epsilon_{SE}$ can be used to certify the validity of the thermodynamic approach; values smaller than 0.05 are obtained when a steady-state modal distribution is reached by the systems.

\section{Experimental setup. Modal losses}\label{sec:MDL}

\begin{figure}[h]
\centering
\includegraphics[width=0.8\textwidth]{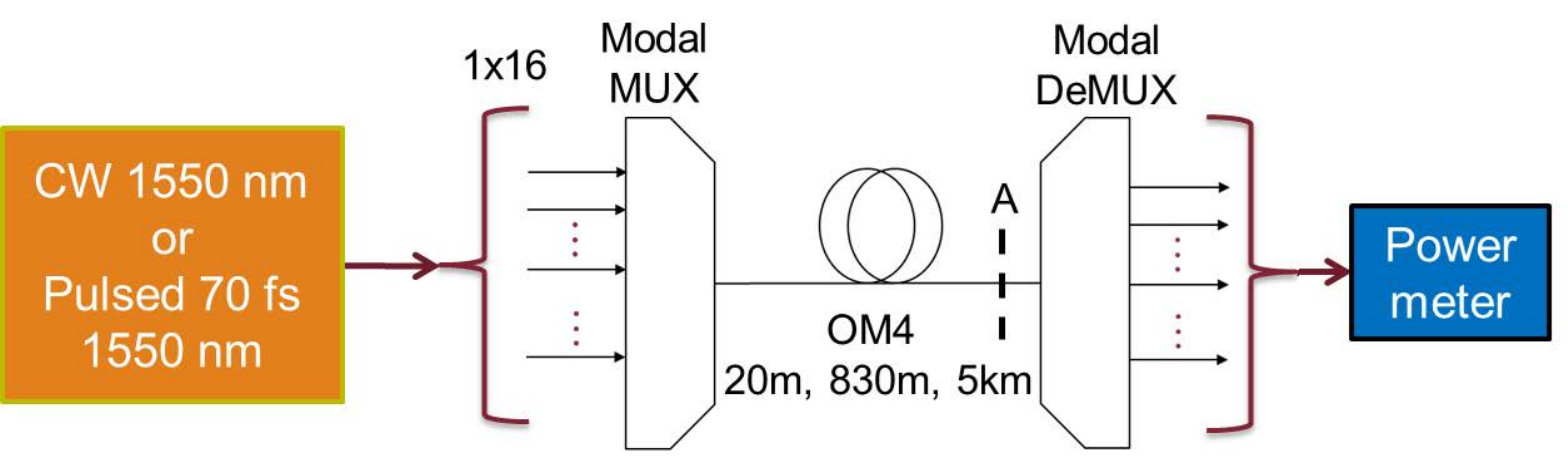}	
\caption{Experimental setup for the characterization of RMC in linear multimode fiber systems.}
\label{fig:Fig1}
\end{figure}

Several experiments, using different modal decomposition methods, sources and power levels have been performed and compared to study the validity of the thermodynamic approach to the RMC-induced modal power redistribution. The main experimental setup is illustrated in Fig. \ref{fig:Fig1}. The source generated 70 fs pulses at wavelength $\lambda=1550$ nm and 100 kHz repetition rate with controllable power; alternatively, a kilohertz bandwidth continuous-wave (CW) source was used at 1550 nm to analyze the role of modal phase coherence. The input signal was split using a $1x16$ splitter whose outputs are 16 single-mode fibers carrying equal pulse energy (with 3\% tolerance); in pulsed regime, pulse energy was controllable between 1 and 100 pJ per mode. Fibers were coupled to one or more inputs of a modal multiplexer (Cailabs Proteus C-15) using multi-plane light conversion (MPLC) technology \cite{Labroille201793}, which coupled the power of each input fiber to one of the first 15 Laguerre-Gauss modes ($LG_{mn}$ or $LG_p$ with $p=1,2,..,15$) of a GRIN OM4 fiber (Thorlabs GIF50E); the fiber was capable of supporting $Q=10$ groups of quasi-degenerate modes at 1550 nm, corresponding to $M=55$ modes per polarization. Different lengths of fiber (20 m, 830 m and 5 km) were spliced to the multiplexer. The fiber output end was spliced to an identical modal demultiplexer to perform the modal decomposition and measure the output modal power distribution. Alternalively, the OM4 bare output end (section $A$) was focused to an infrared camera to measure both the near-field (NF), the far-field (FF) and the total output power. The input power was measured from one of the $1x16$ splitter outputs.

\FloatBarrier

As a preliminary test, it was necessary to isolate the effects of RMC from mode-dependent losses (MDL). The loss coefficient is usually modeled as \cite{Gloge:72} $\alpha=\alpha_0+A(j-1)^2$, with $\alpha_0$ (1/m) the material loss, $A$ (1/m) the modal loss coefficient and $j=1,2,..,Q$ the modal group index.
Material loss was measured coupling power to the fundamental mode $LG_{00}$ and measuring the total output power from 20 m or 5 km of fiber. Comparison provided $\alpha_0=0.192$ (dB/km) $=4.4x10^{-5}$ (1/m) .
Insertion loss $IL_j$ of modal groups was measured by applying power to all modes of a group $j$ and measuring the total power from all modes of the same group using the modal demultiplexer after 20 m of fiber, assuming negligible RMC and modal loss over the short distance. The values in the second column of Tab. \ref{tab1} were obtained; the third column is the group transmittance $T_j(L)=IL_j\exp{(-\alpha_0L)}$ after $L=5$ km, not including modal losses.

Modal loss was measured by applying power to and measuring power $P_j(L)$ from all modes of a group $j$ after 5 km of fiber; third and fourth columns in Tab. \ref{tab1} report the output group power $P_j(L)$ and the mean modal power in the groups $P_j(L)/j$; power levels are well below the values were pulse propagation is affected by the Kerr and Raman nonlinearity; in pulsed regime, 20 pJ pulse energy per mode was used, which corresponds to peak powers hundreds of times the soliton threshold at input, and thousands of times at output. Modal loss power is calculated by normalizing the output power to material and insertion loss as $P_{nj}(L)=P_j(L)/[jT_j(L)]$ (column 6). After normalizing to the power of the fundamental mode, the fractional modal loss $MDL_j$ is obtained on column 7, showing a quadratic reduction of power up to 25\% for modal group $j=5$. A numerical fit is calculated on column 8 according to the theoretical law $\exp[-A(j-1)^2L]$ (nonlinear least squares), obtaining an estimate of the modal loss coefficient $A=4.44x10^{-6}$ (1/m) with fit R-square of 0.88.

\begin{table}[h]
\caption{}
\label{tab1}  
\centering	
\footnotesize
\begin{tabular}{@{}llllllll@{}}
\toprule
$j$ & $IL_j$ & $T_j(L)$ & $P_j(L)$ & $P_j(L)/j$ & $P_{nj}(L)$ & $MDL_j$ & Fit\\
\midrule
- & - & - & [$\mu W$] & [$\mu W$] & [$\mu W$] & - & - \\
\midrule
1 & 0.335 & 0.268 & 1.080 & 1.080 & 4.026 & 1.000 & 1.000\\
2 & 0.326 & 0.261 & 2.082 & 1.041 & 3.983 & 0.989 & 0.978\\
3 & 0.274 & 0.219 & 2.300 & 0.767 & 3.493 & 0.868 & 0.915\\
4 & 0.274 & 0.220 & 2.733 & 0.683 & 3.110 & 0.773 & 0.819\\
5 & 0.249 & 0.200 & 2.996 & 0.599 & 3.001 & 0.746 & 0.701\\
\bottomrule
\end{tabular}
\end{table}

\section{Near and Far-Fields}\label{sec:FF}

\begin{figure}[h]
\includegraphics[width=0.8\textwidth]{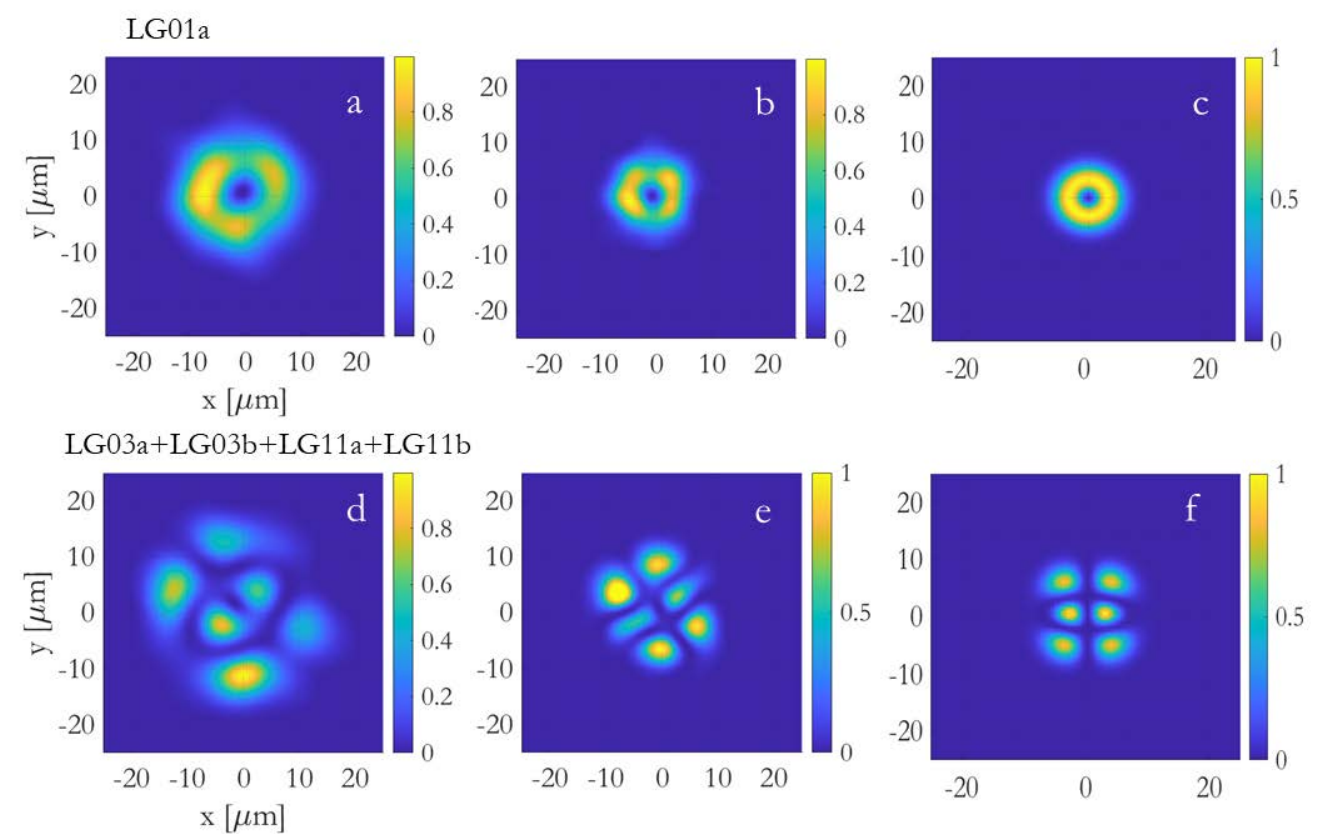}	
\centering	
\caption{(a) Measured NF, (b) measured FF and (c) numerically calculated FF after 10 m of GRIN fiber, when a mode $LG_{01}$ is coupled at input using low power pulses. (d) Measured NF, (e) measured FF and (f) numerically calculated FF after coupling modes $LG_{03a}$, $LG_{03b}$, $LG_{11a}$ and $LG_{011b}$ with same power at input.}
\label{fig:Fig2_1}
\end{figure}

As a second test, the output demultiplexer was removed and the NF and FF were collected using a calibrated IR camera. Figs. \ref{fig:Fig2_1}a and \ref{fig:Fig2_1}b show the measured NF and FF, respectively, after 10 m of fiber (other 10 m remained connected to the output demultiplexer) when the input mode $LG_{01}$ (group $j=2$) is coupled in pulsed regime; Fig. \ref{fig:Fig2_1}c is the numerically calculated FF showing good correspondence.

Figs. \ref{fig:Fig2_1}d,e,f illustrate the measured NF, FF and the numerical FF, respectively, when all modes of group $j=4$ were coupled with same power ($LG_{03a}$, $LG_{03b}$, $LG_{11a}$ and $LG_{011b}$). The numerical coherent sum of the modal FFs matches the measured FF.

\begin{figure}[h]
\includegraphics[width=0.8\textwidth]{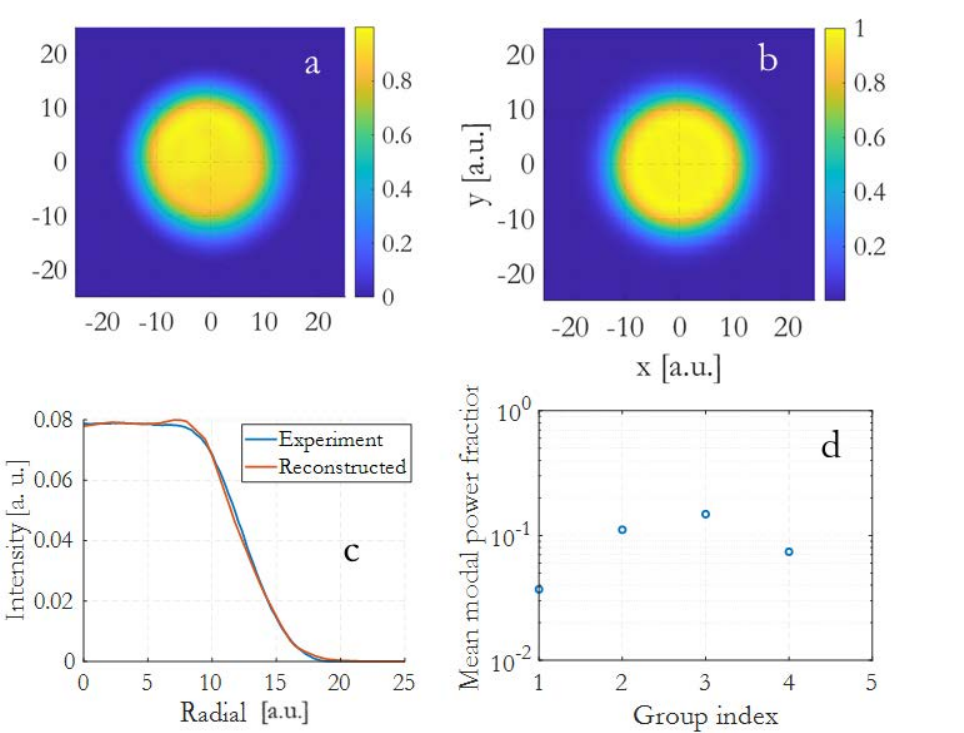}	
\centering
\caption{(a) Measured FF, (b) numerically reconstructed FF and (c) experimental and reconstructed radial intensity profile, after 5 km of GRIN fiber, when coupling modes $LG_{03a}$, $LG_{03b}$, $LG_{11a}$ and $LG_{011b}$ with same power at input. (d) Modal decomposition resulting from the numerical reconstruction.}
\label{fig:Fig2_2}
\end{figure}

When repeating the test after 5 km of fiber, it was detected the effect of modal phase incoherence. Input 70-fs pulses broaden up to 5.5 ns as a consequence of chromatic and modal dispersion; pulses corresponding to different modal group are still time overlapped, however the cumulated phase chirp is so high that mode powers add incoherently at the output. This is shown in the measured FF of Fig. \ref{fig:Fig2_2}a when all modes of group $j=4$ were coupled at the input; the FF appears considerably different from Fig. \ref{fig:Fig2_1}e because of modal de-coherence. A numerical modal decomposition was performed by adding modal powers incoherently and comparing reconstructed and measured FF; to simplify the iteration, it was assumed that all quasi-degenerate modes in a modal group statistically carry the same power at the output. Intensity of the first 15 modes were iteratively weighted by a modal power $P_p$, $p=1,2,..,15$, and added incoherently; the mean radial intensity profile was calculated and compared to the experiment; the best reconstruction in Fig. \ref{fig:Fig2_2}c produced the FF of Fig. \ref{fig:Fig2_2}b and the output distribution of Fig. \ref{fig:Fig2_2}d; this is expressed as the mean modal power fraction $\lvert f_j \rvert^2=\sum_{p} P_p /(j P_{tot})$, being the sum limited to the quasi-degenerate modes of group $j$, $p=j(j+1)/2-j+1,..,j(j+1)/2$. From Fig. \ref{fig:Fig2_2}d, we note that RMC generates an asymmetric flow of power towards the lower-order modes (LOM) and higher-order modes (HOM), which promotes the LOMs.

\section{Modal Distributions by Groups}\label{sec:MD by groups}

A more accurate modal decomposition was obtained by splicing the modal demultiplexer at the output of 830 m or 5 km of fiber. Pulses were coupled with same power, corresponding to 20 pJ energy per pulse per mode, applied to all the modes of a single group $j$ or to all the input modes. 
Output power $P_p$ was measured from the 15 modes, calculating the mean and standard deviations over 60 s acquisitions; the output mean modal power fraction was calculated dividing the insertion loss $IL_j$ and the modal loss $MDL_j$ and respecting the condition $\sum_{j=1}^5 j\lvert f_j \rvert^2=1$ 

\begin{equation}
\lvert f_j \rvert^2=\frac{1}{j IL_j \exp{[-A(j-1)^2L]} P_{tot}} \sum_{p=j(j+1)/2-j+1}^{j(j+1)/2}P_p .
\label{eq:MMPF}
\end{equation}

Power fractions calculated from Eq. \ref{eq:MMPF} are only affected by the RMC.

\begin{figure}[h]
\includegraphics[width=0.8\textwidth]{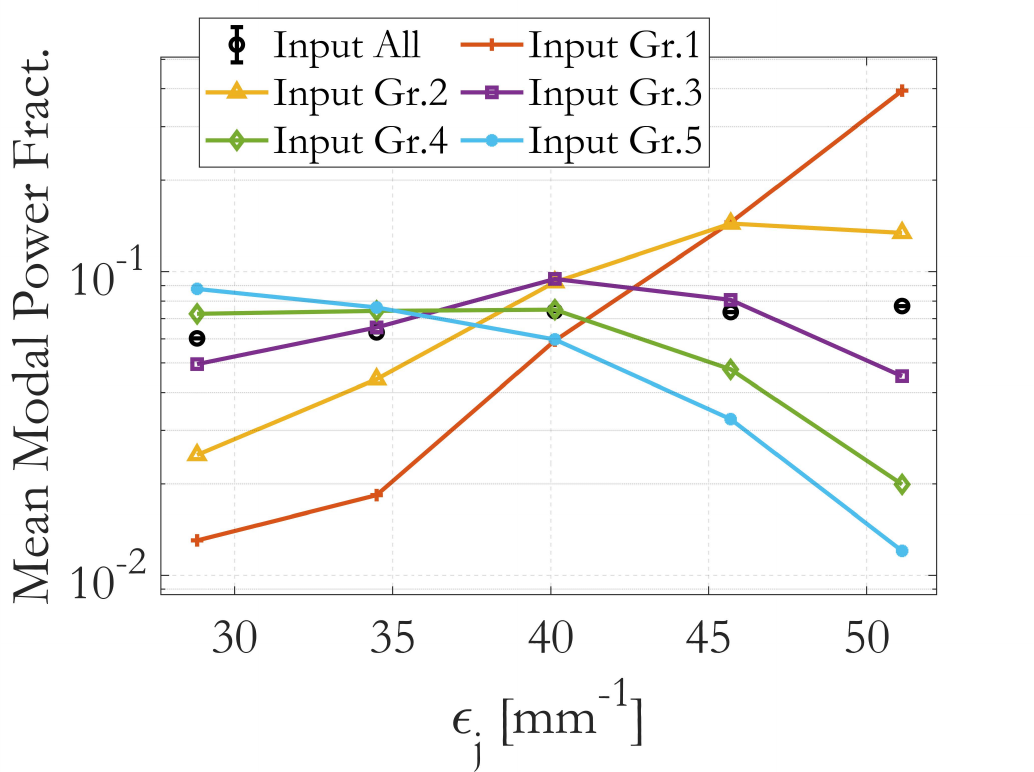}	
\centering	
\caption{Experimental modal decomposition in linear regime after 5 km of GRIN, when input pulses are applied with same power to the modes of the single groups (Input Gr.$j$) or to all input modes (Input All).}
\label{fig:Fig3}
\end{figure}

Figure \ref{fig:Fig3} reports the mean modal power fractions measured against the modal eigenvalues $\epsilon_j$ after 5 km of GRIN, when input pulses are applied to the modes of the single groups (Input Gr.$j$) or to all input modes (Input All). Groups 1 to 5 have differential eigenvalues ordered from the highest to the lowest; it is $\epsilon_Q=0$ for $Q=10$, corresponding to the number of groups supported by the GRIN fiber.

As a general result, when power was coupled to group $j$ at input, the RMC-induced cross-talk was not limited to the single group; group $j-1$ always attracted larger output power respect to $j+1$, indicating that RMC produces an asymmetric flow of power among modal groups.
When power was uniformly coupled to all modes at input, output distribution was not uniform as well and the LOMs were promoted; error-bars are reported in the figure as the standard deviation of the collected output power; negligible errors (less than 0.5\%) indicate a high stability of the output modal distribution in pulsed regime. 

\begin{figure}[h]
\includegraphics[width=0.8\textwidth]{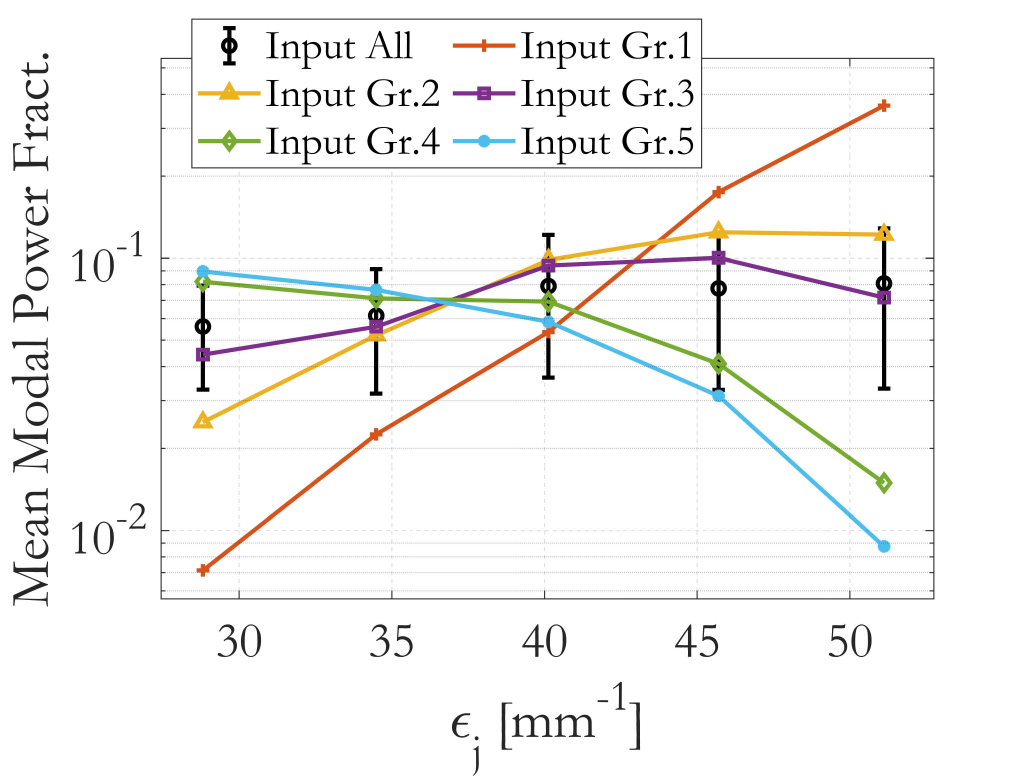}	
\centering	
\caption{Experimental modal decomposition in linear regime after 5 km of GRIN, when CW signals are applied with same power to the modes of the single groups (Input Gr.$j$) or to all input modes (Input All).}
\label{fig:Fig4}
\end{figure}

The experiment was repeated in Fig. \ref{fig:Fig4} after 5 km of fiber, using at input a highly coherent 1550 nm CW source with 10 kHz linewidth (Thorlabs TLX1); the average power applied to each mode at fiber input was 90 $\mu$W. Similar output distributions to the pulsed regime were obtained; however, output power standard deviations were not negligible (up to 50\% of the measured power). The same modal instability was observed in the measured near fields in CW regime, which showed power fluctuations among modes on a time scale of few seconds. A possible explanation to this effect is the partial coherence conserved by the modes at the output, which still add coherently; when modes gather a phase noise induced by the large cumulated chromatic and polarization dispersion, a large instability is observed in the decomposed mode powers. Such effect is not observed when input pulses are used, which are characterized by large bandwidth (few nm).

\section{Modal Distributions for Uniform Input}\label{sec:MD uniform}

\begin{figure}[h]
\includegraphics[width=0.8\textwidth]{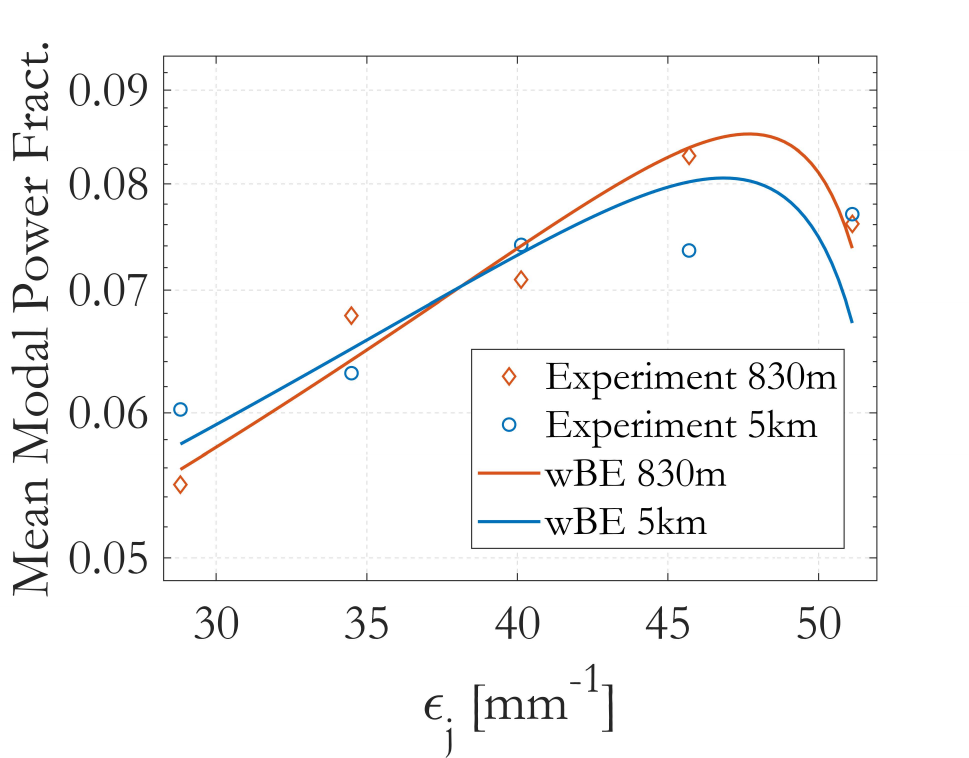}	
\centering	
\caption{Output mean modal power fractions $\lvert f_j \rvert^2$ vs. $\epsilon_j$ measured after 830 m or 5 km of GRIN fiber, when equal energy pulses (20 pJ) are coupled to the 15 input modes. Powers are normalized to the insertion and modal lossess, according to Eq. \ref{eq:MMPF}. The wBE fits are plotted for both cases.}
\label{fig:Fig5}
\end{figure}

The experiment in pulsed regime of Fig. \ref{fig:Fig3} was repeated at 830 m distance. Fig. \ref{fig:Fig5} reports the output mean modal power fractions $\lvert f_j \rvert^2$ vs. $\epsilon_j$ after 830 m and 5 km, when all 15 input modes are coupled with same power, corresponding to a pulse energy of 20 pJ per mode. Power fractions are calculated by Eq. \ref{eq:MMPF}, respecting the condition $\sum_{j=1}^5 j\lvert f_j \rvert^2=1$, such that only RMC affects the plotted distributions.

One of the main results visible in the figure is that the RMC-induced steady-state is characterized by a larger mean power fraction of the fundamental mode. At intermediate distance (830 m), the promotion of the LOMs is clear but not complete. 

In order to explain the results of Fig. \ref{fig:Fig5}, the role of material loss must be taken into account, which is negligible at 830 m (0.2-0.3 dB) and is large after 5 km.
The thermodynamic approach holds when a single experiment conserves the total number of photons $N$; after 830 m distance, the material loss is responsible for a power drop of less than 4\%, which increases to 20\% after 5 km. Hence, when fitting the experimental data using the wBE law of Eq. \ref{eq:BE}, a better result is obtained at 830 m respect to 5 km. The obtained thermodynamic parameters at 830 m (5 km) are: $T=321170$ (310430) (1/m), $\mu'=-67730$ (-72130) (1/m), $\gamma=250$ (210); the error on the state equation $\epsilon_{SE}=8.4 x 10^{-3}$ ($9.6 x 10^{-3}$). The fit R-squares are 0.93 after 830 m, and 0.62 after 5 km, confirming that negligible losses are a condition for the validity of the thermodynamic approach, while dispersion-induced pulse broadening does not invalidate the thermodynamics in the linear regime.

Fits confirm that the wBE law is able to describe the output modal distribution affected by RMC; a very good correspondence is obtained when material losses are limited to less than 0.5 dB.

\section{Power-flow Simulations}\label{sec:Power flow}

\begin{figure}[h]
\includegraphics[width=0.8\textwidth]{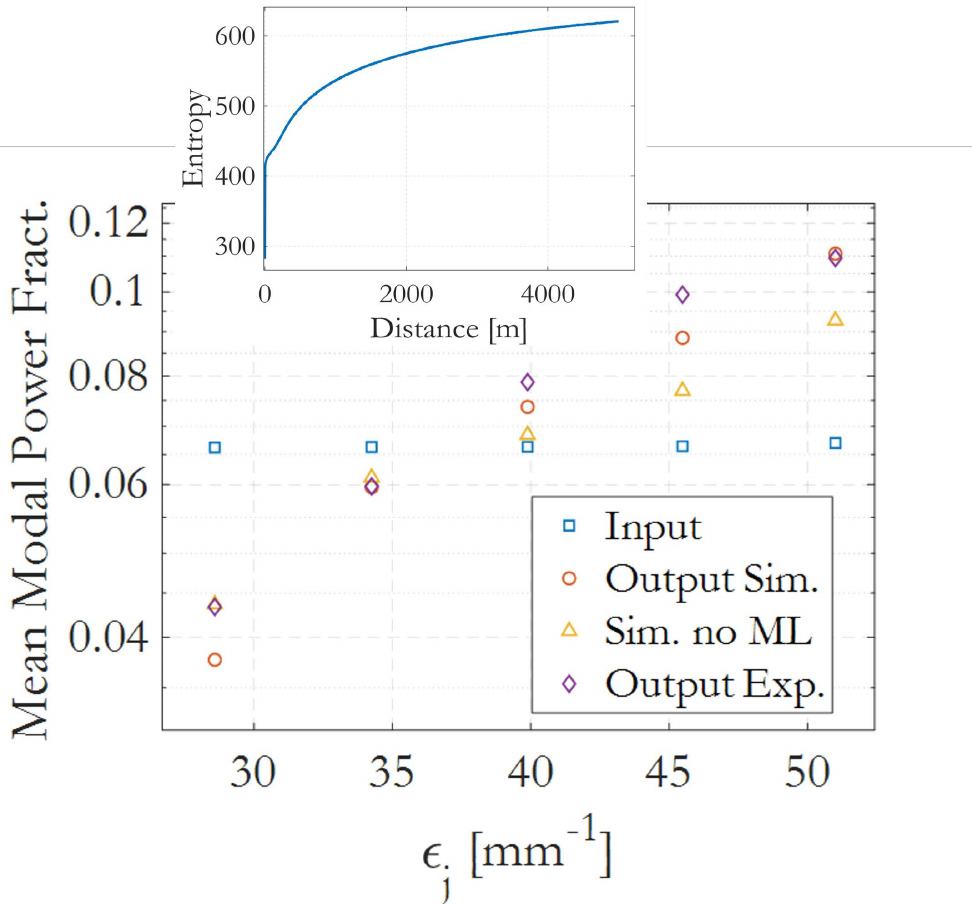}	
\centering	
\caption{Power-flow simulations of the experiment  with 5 km of GRIN fiber and 15 modes with uniform input power. (Input) Input mean modal power fraction vs. modal group eigenvalues. (Output Sim.) Simulated output distribution, including modal losses. (Sim. no ML) Simulated output distribution with no modal losses. (Output Exp.) Experimental output distribution.}
\label{fig:Fig6}
\end{figure}

Power-flow simulations were performed in Fig. \ref{fig:Fig6} using the model from \cite{Savovi2019PowerFI}, for the case of 5 km of GRIN fiber and 15 modes coupled at input with uniform power (Input). Red circles (Output Sim.) provide the obtained numerical output distribution using a RMC coupling coefficient $D=8.5x10^{-5}$ $m^{-1}$ and the measured modal loss coefficient $A=4.44x10^{-6}$ $m^{-1}$, in agreement with the experimental results (Output Exp.). Yellow triangles (Sim. No ML) provide the simulated output distribution without modal loss ($A=0$), showing that RMC alone can promote the lower-order modes at steady state, producing similar results to MDL. 

In the experiments, it was therefore important to study the output distribution after isolating the two processes, following the approach of Eq. \ref{eq:MMPF}.
Experiments in this work confirm the prediction of the theory in \cite{Gloge:72,Savovi2019PowerFI}, stating that mode-coupling coefficients are not equal for power flow towards lower-order or higher-order groups. It results a net power flow towards the LOMs.

The inset in Fig. \ref{fig:Fig6} further shows the entropy evolution with distance, calculated using Eq. \ref{eq:EntropyB} on the simulated data including modal losses and using power levels corresponding to the 20 pJ propagated pulses. Hence, the power-flow model confirms the predictions by the thermodynamic theory, providing modal distributions which evolve to a steady-state corresponding to a maximum of Boltzmann entropy.

\section{Modal Distributions at Quantum Power}\label{sec:MD power}

\begin{figure}[h]
\includegraphics[width=0.8\textwidth]{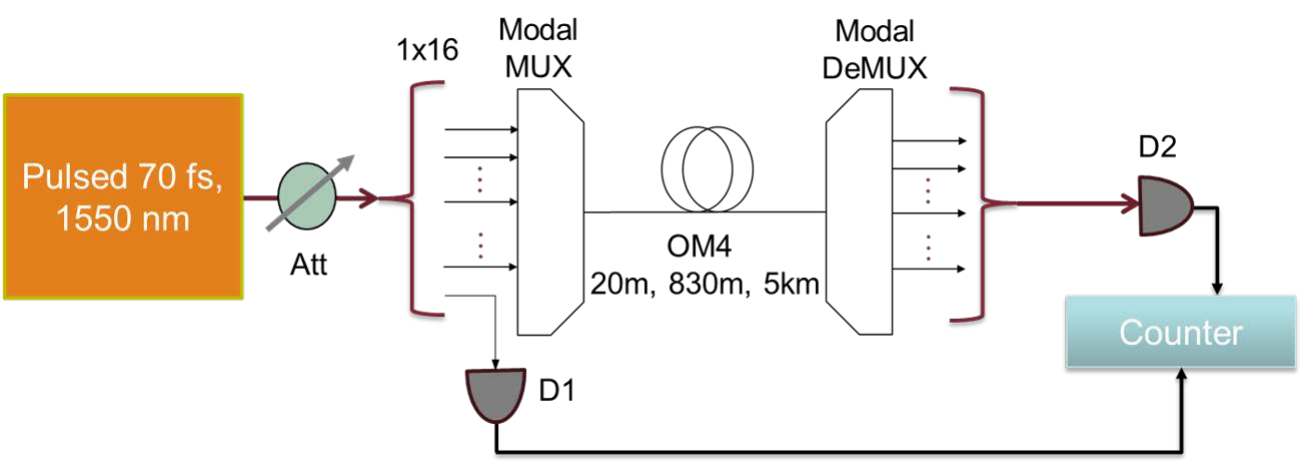}	
\centering	
\caption{Experimental setup for the characterization of RMC in quantum regime.}
\label{fig:Fig7}
\end{figure}

Another aspect worthy of attention is testing the ergodicity of the RMC process, when it is evaluated in thermodynamic terms. The experimental setup was modified like in Fig. \ref{fig:Fig7}. Pulses at 1550 nm and 100 kHz repetition rate were attenuated to quantum level (1 photon per pulse per mode), and injected into the first 15 LG modes though a 1x16 splitter and the modal Mux. One of the splitter output was connected to a single-photon detector $D1$ (IDQube-NIR-FR) with sub-microsecond dead-time, 10\% efficiency and operated in free-running mode. A second detector $D2$ was connected to one of the output modes after the modal DeMux. Detector counts were collected by a time controller (ID1000-MASTER) with picosecond resolution. 

Both $D1$ and $D2$ could detect a maximum of one photon per pulse due to the larger dead-time with respect to the output pulsewidth (5.5 ns); hence, 100 kHz detection rate on $D1$ would correspond to more than 10 photons per pulse, after the 10\% detection efficiency. Power at the input of each mode was regulated to have a detector rate of 10 kHz on $D1$, corresponding to an average of 1 photon per pulse per mode. Counts were collected on $D1$ and $D2$ on a 30 s integration time. 

\begin{figure}[h]
\includegraphics[width=0.8\textwidth]{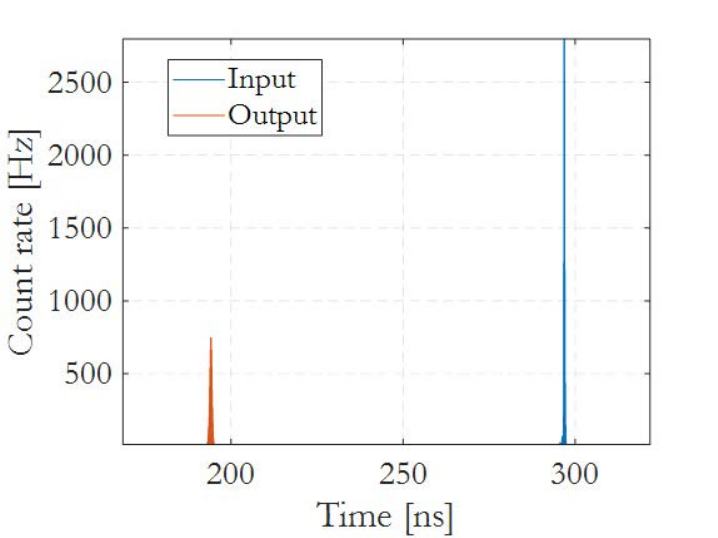}	
\centering	
\caption{Photon-count histogram example for one fiber mode at the input and output of the line.}
\label{fig:Fig8}
\end{figure}

\begin{figure}[h]
\includegraphics[width=0.8\textwidth]{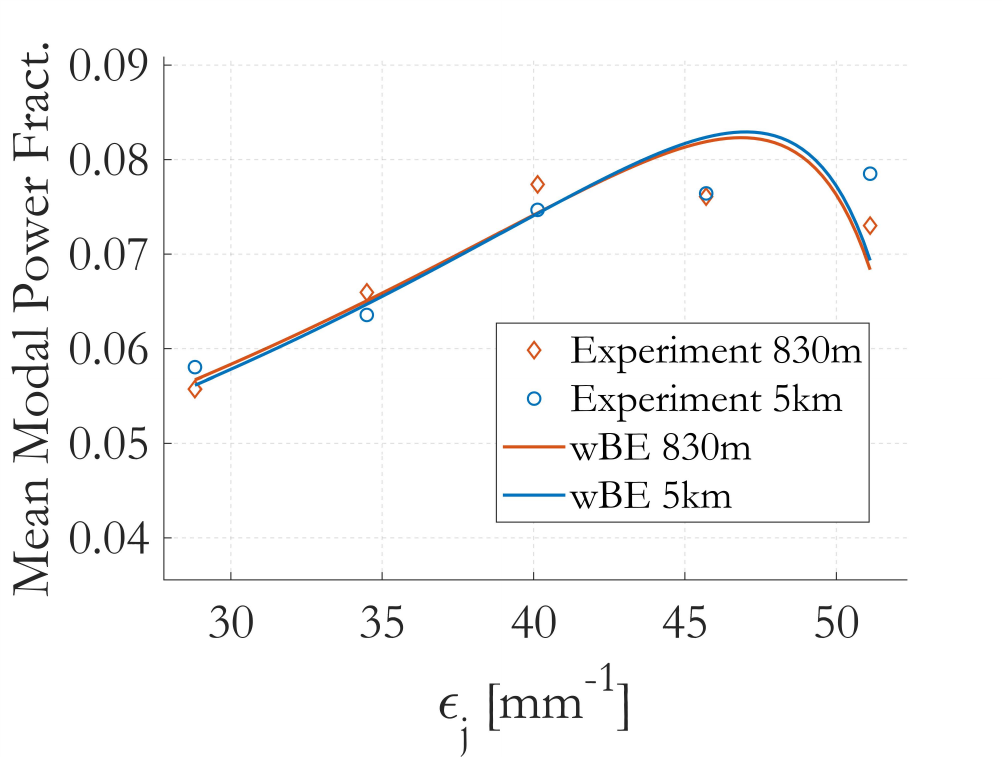}	
\centering	
\caption{Output mean modal power fractions $\lvert f_j \rvert^2$ vs. $\epsilon_j$ measured after 830 m or 5 km of GRIN fiber, when single-photon pulses are coupled to the 15 input modes. Powers are measured as photon counts, and are normalized to the insertion and modal lossess. The wBE fits are plotted for both cases.}
\label{fig:Fig9}
\end{figure}

Figure \ref{fig:Fig8} is an example of the histograms from the input and output detectors, with arbitrary time delay and recording time. By integrating them on a narrow time window (100 ns), it was possible to efficiently eliminate the detectors random noise. By comparing the output and input counts, and normalizing to the insertion and modal loss using Eq. \ref{eq:MMPF}, where power is replaced by the counts, it was possible to extract the mean modal power fractions of Fig. \ref{fig:Fig9}. The ergodicity of the RMC process allows to construct the output power distributions from a large number of single-photon pulses repeated in time, quantified in 15 modes times 30 seconds times $1x10^5$ repetition rate, or $N=4.5x10^7$. 
When fitting the experimental data using the wBE law, Eq. \ref{eq:BE}, the thermodynamic parameters at 830 m (5 km) are: $T=83924$ (72679) (1/m), $\mu'=-75311$ (-75456) (1/m), $\gamma=42.8$ (35.6); the error on the state equation $\epsilon_{SE}=8.2 x 10^{-3}$ ($1.2 x 10^{-2}$). The fit R-squares are 0.892 after 830 m, and 0.802 after 5 km. Hence, the wBE is able to describe the output modal distribution also when uniform quantum power is coupled at input, provided material losses are negligible.

\section{Discussion and Conclusion}

Several experiments, aimed to characterize the thermodynamics of RMC at telecom wavelength (1550 nm), were repeated using a CW kilohertz-bandwidth source, or 70-fs terahertz-bandwidth input pulses. Output modal decomposition were performed after 830 m and 5 km of GRIN fiber by reconstruction of the output FF or using an output modal demultiplexer based on MPLC technology. In pulsed linear regime, pulse energy was limited to 20 pJ per mode at the MM fiber input, in order to work at power levels which are hundreds to thousand times lower than the nonlinear soliton regime (whose peak power in MMF is of the order of 30 kW \cite{Zitelli:JosaB24}). In CW linear regime, modal power was limited to 90 $\mu$W for the same reason. In all experiments, the measured modal and insertion loss were normalized to isolate the effects of RMC.

Comparable output distributions, measured as mean modal power fractions $\lvert f_j \rvert^2$ vs. $\epsilon_j$, were obtained from experiments using the different sources and modal decomposition methods, both after 830 m and 5 km of fiber. When single modal groups were coupled, the output distribution denoted an asymmetric diffusion process which promoted the lowest-order groups, as confirmed by the power-flow simulations. When a uniform modal distribution was coupled at input, the output distribution converged to a steady-state which is accurately described by the wBE law of Eq. \ref{eq:BE}, in particular when linear loss is negligible.

The use of CW beams with large coherence revealed instability in the power of the decomposed modes, due to the partial phase coherence preserved by the modes despite the accumulated chromatic dispersion. The power fluctuations totally disappeared when using large bandwidth pulses.

Experiments were repeated using single-photon pulses applied to the input modes, and replacing the power measurements with photon-count techniques. The ergodicity of the RMC process was demonstrated by observing that comparable output distributions could be measured using 20 pJ pulses carrying $1.5x10^8$ photons or single-photon pulses repeated $N=4.5x10^7$ times. When uniform distribution of photon counts was applied at the input, the output distribution was described by a wBE with error coefficients comparable to the linear regime. 

The obtained theoretical, numerical and experimental results indicate that RMC affects quantum channels in a SDM systems in a similar way to classical channels. In the quantum case, distribution statistics are gathered from a large number of repeated single-photon pulses. Both in linear and quantum regimes, power distributions at the output of MMFs reach steady-states which can be described in thermodynamic terms by a wBE law. In a first aspect, this demonstrates the inaccuracy of the assumptions of a RMC-induced cross-talk with uniformly distributed weight, or limited to the quasi-degenerate modes of a same group. In a second aspect, the knowledge of a thermodynamic law which describes the output modal power distribution can simplify the design of SDM fiber systems, and eventually MIMO processing techniques in linear as well in quantum transmission systems. In the second case, this observation paves the way to a new discipline that we could define as optical quantum thermodynamics, which allows the study of quantum SDM systems using thermodynamic laws.

\section{Funding}
Project ECS 0000024 Rome Technopole, Funded by the European Union - NextGenerationEU.

\end{document}